\documentclass[preprintnumbers,showpacs,twoside,twocolumn]{revtex4}

\usepackage{latexsym}
\usepackage{graphics}

\preprint{Applied Physics A (2004) Accepted}

\begin{document}

\title{Phosphorus cluster production by laser ablation}
\author{A.V. Bulgakov}
\email{bulgakov@itp.nsc.ru}
\author{O.F. Bobrenok} 
\affiliation{Institute of Thermophysics SB RAS,\\ 1 Lavrentyev Ave., 630090 Novosibirsk, Russia} 
\author{I. Ozerov}
\email{ozerov@crmcn.univ-mrs.fr}
\author{W. Marine}
\author{S. Giorgio}
\affiliation{CRMC-N, UPR 7251 CNRS, Universit\'{e} de la M\'{e}diterran\'{e}e, Facult\'{e} des Sciences de Luminy, Case 901,\\ 13288 Marseille Cedex 9, France}
\author{A. Lassesson}
\author{E.E.B. Campbell}
\affiliation{School of Physics and Engineering Physics,\\ G\"{o}teberg University and Chalmers University of Technology, SE-41296, G\"{o}teberg, Sweden}
%\offprints{}
%\textit{e-mail: bulgakov@itp.nsc.ru}

%\date{Received: date / Revised version: date}
% The correct dates will be entered by the editor
%

\hyphenation{amo-unts}
\begin{abstract}
\label{abstract}
Neutral and charged phosphorus clusters of a wide size range have 
been produced by pulsed laser ablation (PLA) in vacuum at 532, 337, and 193 
nm ablating wavelengths and investigated by time-of-flight mass 
spectrometry. The neutral P$_{n}$ clusters are even-numbered with local 
abundance maxima at $n = 10 $ and 14, while the cationic and anionic clusters 
are preferentially odd-numbered with P$_{7}^{ + }$, P$_{21}^{ + }$, and 
P$_{17}^{ - }$ being the most abundant ions. The dominance of the magic 
clusters is more pronounced at 337-nm ablation that is explained by 
efficient direct ejection of their building blocks under these conditions. 
Nanocrystalline phosphorus films have been produced by PLA in ambient helium 
gas.

%\textbf{PACS:} 52.38.MF; 61.46.+w; 79.20.Ds

\end{abstract}

\maketitle

One of the greatest potentials of pulsed laser ablation (PLA) is in the 
development of novel nanoscale materials. Recently there has been renewed 
interest in the study of phosphorus clusters and nanostructures as potential 
candidates to form fullerene-like and nanotubular materials \cite{1,2,3,4}. Elemental 
phosphorus has been obtained in more allotropic modifications than any other 
element \cite{5} and one would thus expect a variety of structural forms of 
P$_{n}$ clusters. Phosphorus clusters have been studied extensively in the 
last decade by theoretical approaches focused mainly on two structural 
families, cages (polyhedra) and chains \cite{3,4,5,6,7,8,9,10,11,12,13}. A number of structures for 
P$_{n}$ clusters ($n > 4$) have been proposed as energetically more stable 
than tetrahedral P$_{4}$. Also, the viability of phosphorus nanotubes has 
been predicted \cite{1}.

The theoretical suggestions, however, are still not confirmed by experiment. 
The observed dominance of P$_{4}$ in phosphorus vapor and apparent 
instability of larger clusters provided a puzzle for several decades \cite{6}. 
Only a few experiments have been performed on clusters larger than P$_{4}$ 
\cite{9,11,14,15}. Recently, a wide spectrum of neutral and cationic phosphorus 
clusters were synthesized by visible PLA \cite{2}. Different cluster 
distributions were observed under far-UV PLA resulting in stable hydride 
phosphorus clusters \cite{16}.

In this work we continue our investigations on phosphorus cluster production 
by PLA. Three different ablation regimes, at 532 nm, 337 nm, and 193 nm 
laser wavelengths, have been studied and compared with respect to P$_{n}$ 
cluster generation. Mechanisms of cluster formation under PLA conditions are 
discussed. In addition, the first attempt to produce nanocluster phosphorus 
films by PLA technique has been performed.

\section{Experiment}

The apparatus used for laser ablation and cluster production and detection 
was described earlier \cite{2,16,17}. The target (crystalline red phosphorus of 
99.999{\%} purity with respect to metals) was placed in a rotating holder in 
a vacuum chamber (base pressure 10$^{ - 5}$ Pa) and irradiated by a ns laser 
pulse. Three different laser systems operating at 337 nm (10 ns pulse, 
N$_{2}$ laser), 532 nm (13 ns, Nd:YAG laser, 2nd harmonic), and 193 nm (15 
ns, ArF laser) were used for ablation. The laser fluence at the target was 
varied in the range 20 -- 800 mJ/cm$^{2}$ for each wavelength. The relative 
abundance of neutral and charged particles in the PLA plume was analyzed by 
time-of-flight (TOF) mass spectrometry. The plume expanded under field-free 
conditions over a distance of 6 cm (at 532-nm ablation) or 7 cm (at 337 and 
193-nm ablations) towards a repeller grid where the plume ions were sampled 
by pulsing the grid at a time delay $t$ after the laser pulse. Electron-impact 
ionization (90 eV) and a plasma suppressor were used to investigate neutrals 
\cite{2}. Every mass spectrum was summed over 200 laser shots.

The nanostructured phosphorus films were deposited on fused silica 
substrates using 193-nm PLA both in vacuum and in an ambient gas. In the 
latter case, a continuous flux of pure helium was introduced into the 
chamber at a pressure of 400 Pa. The ArF laser fluence was set at 650 
mJ/cm$^{2}$. The substrates were placed at a distance of 5 cm from the 
target and kept at either room or liquid nitrogen (LN) temperature. The 
films were characterized by transmission electron microscopy (TEM) and by 
electron diffraction (ED) using a Jeol 2000FX microscope.

\section{Results and Discussion}

\begin{figure}[!]
\resizebox{0.48\textwidth}{!}{\includegraphics{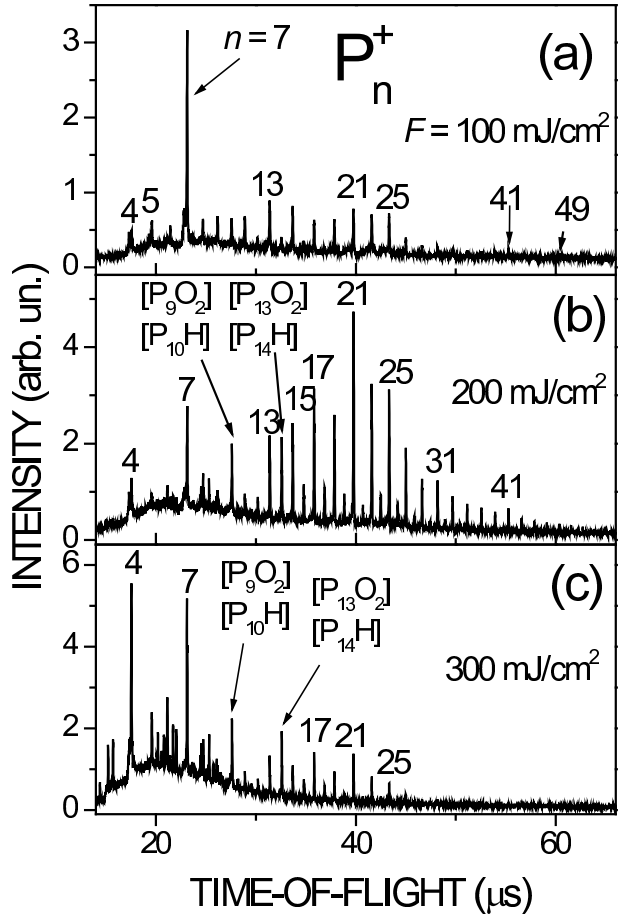}} \caption{\label{fig1}
Mass spectra of cationic phosphorus clusters produced by 
337-nm laser ablation at time delay $t$ = 62 $\mu $s for three different laser 
fluences.}
\end{figure}

With the 532 nm and 337 nm ablating wavelengths, neutral phosphorus clusters 
P$_{n}$ of a wide size range (up to $n = 40$), their cations P$_{n}^{ + }$ (up 
to $n = 91$), and anions P$_{n}^{ - }$ (up to $n = 49$) have been produced in 
abundance in the PLA plume. Figure \ref{fig1} shows the TOF mass spectra of the 
cationic species at 337-nm ablation for three different laser fluences. The 
mass spectra were taken at an "optimum" delay time $t$ corresponding to maximum 
yield of P$_{n}$ clusters with $n > 9$. An interesting feature of the mass 
spectra is the strong fluence dependence for generation of large phosphorus 
clusters that dominate in the plume at a narrow fluence range around 200 
mJ/cm$^{2}$ but represent only minor constituents at 100 and 300 
mJ/cm$^{2}$. In contrast, smaller P$_{n}$ clusters ($n < 9$) are observed in a 
much larger fluence range and their concentration increases progressively 
with fluence.

Another interesting feature of the cation mass spectra shown in Fig. \ref{fig1} is the 
distinct odd-even alternation with domination of odd-numbered clusters and 
local maxima at $n = 17, 21, 25, 31, 41,$ and 49. Under the "optimum" 
conditions, P$_{21}^{ + }$ is remarkably prominent with its peak intensity 
at least a factor of 2 larger than that of other clusters (Fig. \ref{fig1}b). It 
should be noted here that direct comparison of intensities of small and 
large clusters is difficult because of the inherent decline in detector response at high 
masses. For fairly large clusters, the detection efficiency decreases as the 
particle impact velocity decreases down to a threshold value of $\sim $ 20 
km/s \cite{18}. For the present experiments, the ion impact energy was 5 keV and thus 
clusters with masses above roughly 2500 u (i.e., at $n > 80$ for P$_{n}$ 
clusters) were detected with significantly decreasing efficiency. In the 
lower mass range, however, the detector response is a rather weak function 
of impact velocity as was confirmed by a calibration of our microchannel 
plate detector with a C$_{60}$ ion beam. For instance, the detection 
efficiencies for two magic clusters P$_{21}^{ + }$ and P$_{41}^{ + }$ 
differ by a factor of $\sim $ 1.4 whereas their experimental intensities differ by an 
order of magnitude. Thus, the observed abundance distribution is expected
to give a good indication of the true one.

Even-numbered P$_{n}^{ + }$ clusters are present in minor amounts up to 
$n$ = 40 (Fig. \ref{fig1}) as double peaks separated by 1 u. The second, stronger, peak 
in the bunch is attributed to either P$_{n}^{ }$H$^{ + }$ or P$_{n - 
1}^{ }$O$_{2}^{ + }$ ($n$ = even) which could not be distinguished in our 
experiment. Both H and O atoms come from trace impurities in the target as 
was confirmed by laser desorption MS analysis. The relative abundance of the 
compound clusters increases with fluence (Fig. \ref{fig1}c). It has already been 
speculated that a dodecahedral cage-like P$_{20}$, stabilized with an 
additional fourfold coordinated P atom in the laser plasma, could be 
especially stable and explain the magic number of 21 \cite{2}. The energetic 
stability of the dodecahedral structure for neutral P$_{20}$ was recently 
confirmed by calculations \cite{4,13}. A similar cation mass spectrum as shown in 
Fig. \ref{fig1}b was obtained with a 532-nm ablating wavelength at an ``optimum'' 
laser fluence of around 300 mJ/cm$^{2}$ \cite{2}. However, the 337-nm ablation 
resulted in a higher abundance of large P$_{n}^{ + }$ clusters and in a 
more pronounced preference of the P$_{21}^{ + }$ magic cluster. Further 
reduction in laser wavelength down to 193 nm resulted again in a lower 
abundance of bare P$_{n}$ clusters and in generation of compound clusters 
with the P$_{23}^{ }$H$_{6}^{ + }$ cation being the most stable \cite{16}.

\begin{figure}[!]
\begin{center}
\resizebox{0.45\textwidth}{!}{\includegraphics{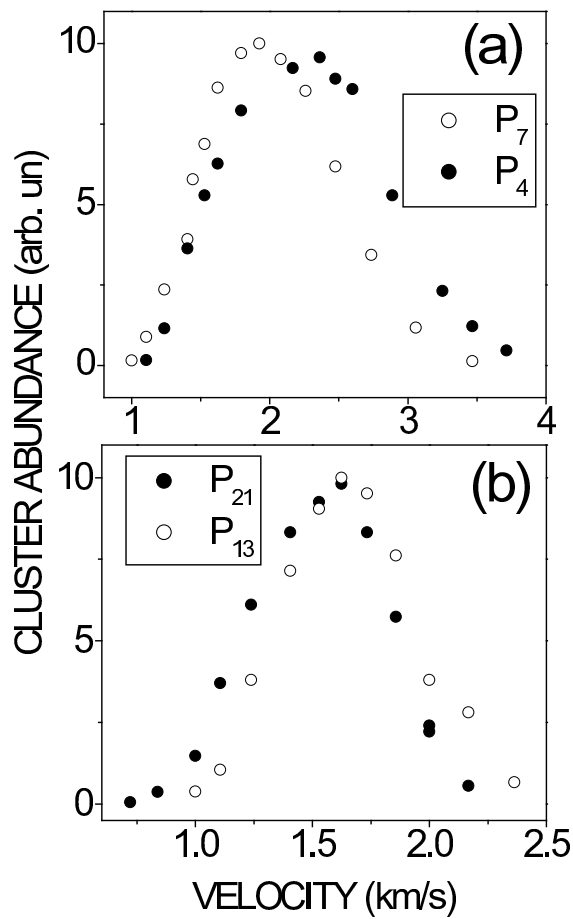}}
\end{center}
\caption{\label{fig2}
Velocity distributions for small (a) and large (b) 
phosphorus cluster cations produced by 532-nm laser ablation at 300 
mJ/cm$^{2}$.}
\end{figure}

To further elucidate the cluster formation process, the plume dynamics has 
been examined by varying the time delay $t$. Figure \ref{fig2} illustrates typical 
velocity distributions for several representative clusters obtained at 
532-nm ablation under "optimum" conditions. A clearly different expansion 
behavior is seen for small ($n < 9$) and larger species. The large P$_{n}$ 
clusters have near equal expansion velocities ($\sim $ 1.6 km/s for these 
conditions, Fig. \ref{fig2}b) over a wide size range in spite of the large difference 
in their masses. This implies a gas-phase condensation mechanism for their 
formation rather than direct ejection \cite{2,16}. Smaller P$_{n}$ clusters have 
size-dependent velocities ($\sim $2 km/s for P$_{7}^{ + }$ and $\sim $2.4 
km/s for P$_{4}^{ + }$, Fig. \ref{fig2}a) and appear to be ejected from the target. 
A plausible explanation for the efficient generation of large phosphorus 
clusters by PLA without a carrier gas is, therefore, that their building 
blocks (small P$_{n}$ clusters) have been already formed in the target and 
ejected thus facilitating the gas-phase cluster growth. It is important that 
the building blocks are larger than P$_{4}$ in order to avoid simple 
polymerization of P$_{4}$ during the clustering process \cite{6}. 

The highest abundance of large P$_{n}^{ + }$ clusters observed under 
337-nm ablation indicates that this regime appears to be around the most 
favorable one for emission of their building blocks via a non-thermal 
mechanism. An additional evidence for the direct cluster ejection is that 
the relatively large species ($n > 4$) were observed at very low threshold 
fluences (around 100 mJ/cm$^{2}$ for 532 nm and 50 mJ/cm$^{2}$ for 337 nm) 
when no smaller fragments were detected. Longer laser wavelengths result in 
a higher contribution of the thermal vaporization mechanism and thus in 
higher abundance of atoms and small P$_{n}$ molecules ($n < 5$) among the 
vaporized products \cite{5,16}. On the other hand, for far-UV laser ablation, 
ejection of the intact building blocks is less efficient due to their 
subsequent photodissociation and more efficient photoionization of the 
target impurities promotes formation of the compound clusters \cite{16}.

\begin{figure}[!]
\resizebox{0.48\textwidth}{!}{\includegraphics{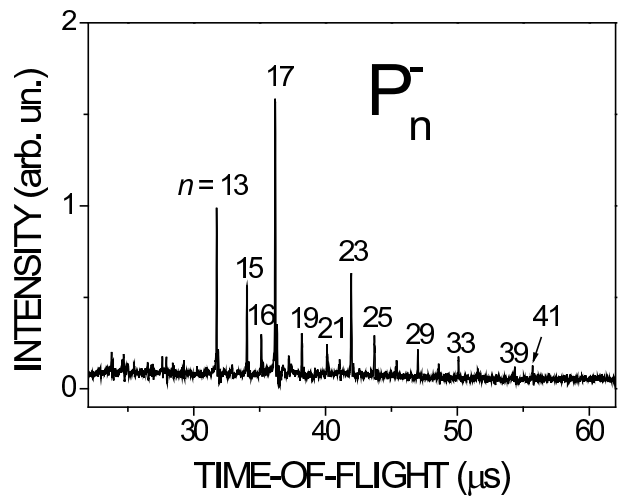}} \caption{\label{fig3}
Mass spectrum of anionic phosphorus clusters produced by 
337-nm laser ablation at $t = 62 \mu $s and $F = 200 $ mJ/cm$^{2}$.}
\end{figure}

Along with the P$_{n}^{ + }$ cations, anionic and neutral phosphorus 
clusters of a wide size range are also abundant in the plume. Figure \ref{fig3} shows 
a mass spectrum of P$_{n}^{ - }$ anions produced under the same conditions 
as for Fig. \ref{fig1}b. Again a pronounced odd-even alternation is observed with 
preferred formation of odd-numbered clusters. However, the magic numbers are 
essentially different from those for P$_{n}^{ + }$ clusters. At $n > 23$, a 
four-fold periodicity for anionic magic numbers is evident. The strongest 
peak corresponds to P$_{17}^{ - }$. At $n > 10$, P$_{16}^{ - }$ and 
P$_{18}^{ - }$ anions are the only observed even-numbered bare clusters. A 
similar anion distribution was observed at 532-nm ablation \cite{15}, again with 
less pronounced magic clusters. The difference in abundance distributions 
for anionic and cationic phosphorus clusters suggests that their stable 
structures are different as well. Indeed, recent calculations for small 
P$_{n}$ clusters (up to $n = 9$) showed that the cluster geometry changed 
considerably when the degree of charging was changed \cite{3}.

In contrast to the ionized species, neutral P$_{n}$ cluster are 
even-numbered \cite{2,16}. The magic clusters are observed at $n = 10$ and 14 that 
correspond to the local maxima for the compound cationic clusters at high 
fluences (Fig \ref{fig1}c). A local abundance minimum at $n = 22$ indicates the presence 
of polyhedral P$_{n}$ in the PLA plume \cite{2}. The comparison of the absolute 
yields for neutral and charged clusters is not straightforward since the 
electron impact ionization efficiency for P$_{n}$ clusters is unknown. 
Assuming the average cluster ionization cross-section to be 2 $\times $ 
10$^{ - 15}$ cm$^{2}$ (maximum value for P$_{4}$ \cite{19}), we have found that 
the concentration of neutral P$_{n}$ clusters is at least an order of 
magnitude higher than for the corresponding P$_{n + 1}$ cations. 
Furthermore, the distribution of the neutrals could be distorted by 
ionization-induced fragmentation. We expect, however, that this has only a 
minor effect in our case since, at least for small P$_{n}$ clusters, 
dissociative ionization is a considerably less efficient process than 
electron impact direct ionization \cite{19}.

\begin{figure}[!]
\resizebox{0.48\textwidth}{!}{\includegraphics{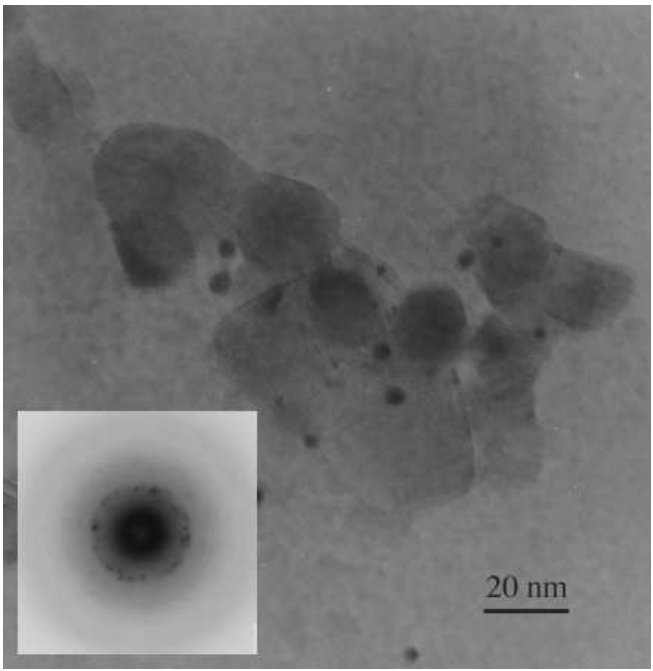}} \caption{\label{fig4}
TEM image of nanocrystalline phosphorus film produced by 
193-nm laser ablation in 400 Pa helium. The inset shows the corresponding ED 
pattern.}
\end{figure}

In spite of the relatively efficient generation of P$_{n}$ clusters under 
PLA in vacuum, the produced clusters are still fairly small and present in 
the plume in a marginal quantity. In order to synthesize a large quantity of 
nanoclusters we performed PLA in a helium atmosphere. The role of the 
ambient gas is to confine the laser ablation plume thus favoring gas-phase 
condensation during the collisional stage of plume expansion. The initial 
small clusters act as nucleation centers for the cluster growth. Figure \ref{fig4} 
shows a TEM image of a sample deposited on a substrate at LN temperature. 
The film consists of near-spherical clusters with an average size of about 
20 nm and has an intense orange color. Some smaller clusters with mean size 
of 3 nm are also present in the film. The electron diffraction pattern 
(inset in Fig. \ref{fig4}) corresponds to a crystalline material with the spacing 
between atomic planes of 2.8 $\pm $ 0.15, 2.1 and 1.8 {\AA}. The complex 
diffraction pattern corresponds to the coexistence of several different 
structures of red phosphorus \cite{5}. Films deposited at room temperature are 
composed of nanocrystals with larger sizes, probably due to their 
aggregation and coalescence on the substrate. In contrast to the crystalline 
films obtained in the helium atmosphere, nanostructured films deposited in 
vacuum from small phosphorus clusters were amorphous. We suggest that the 
high kinetic energy of the species arriving on the substrate results in 
cluster melting followed by a rapid cooling without crystallization.

\section{Conclusions}

Neutral and charged phosphorus clusters have been produced by visible (532 
nm) and near-UV (337 nm) PLA under vacuum conditions in narrow laser fluence 
ranges centered at around 200 and 300 mJ/cm$^{2}$ for 337 and 532 nm, 
respectively. A series of magic clusters, particularly neutral P$_{10}$ and 
P$_{14}$, cationic P$_{7}^{ + }$ and P$_{21}^{ + }$, and anionic 
P$_{17}^{ - }$ have been observed. Preferences of the magic clusters are 
more pronounced at 337-nm PLA that is explained by efficient direct ejection 
of their building blocks (small P$_{n}$ clusters with $n < 9$) under these 
conditions. Nanocrystalline phosphorus films have been deposited by PLA 
in an ambient gas atmosphere. The crystallites have a mean size of about 20 
nm and their crystalline structure corresponds to that of red phosphorus.

\acknowledgements
{This work was partly supported by the Russian Foundation for Basic Research 
(Grant No. 02-03-32221a) and the Royal Swedish Academy of Science (KVA).}

\end{document}